\documentclass[conference]{IEEEtran}
\IEEEoverridecommandlockouts

\usepackage{cite}
\usepackage{amsmath,amssymb}
\usepackage{graphicx}
\usepackage{url}
\usepackage{booktabs}
\usepackage[ruled,vlined,linesnumbered]{algorithm2e}

\begin{document}

\title{An SLO Driven and Cost-Aware Autoscaling Framework for Kubernetes}

\author{
Vinoth Punniyamoorthy$^{1}$,
Bikesh Kumar$^{2}$,
Sumit Saha$^{3}$,
Lokesh Butra$^{4}$,\\
Mayilsamy Palanigounder$^{5}$,
Akash Kumar Agarwal$^{6}$,
Kabilan Kannan$^{7}$\\[1ex]

\small $^{1,2,7}$ IEEE Senior Member, USA\\
\small $^{3}$ East West Bank, USA\\
\small $^{4,5}$ NTT Data, USA\\
\small $^{6}$ Albertsons, USA
}

\maketitle

\begin{abstract}
Kubernetes provides native autoscaling mechanisms, including the Horizontal Pod
Autoscaler (HPA), Vertical Pod Autoscaler (VPA), and node-level autoscalers, to
enable elastic resource management for cloud-native applications. However,
production environments frequently experience Service Level Objective (SLO)
violations and cost inefficiencies due to reactive scaling behavior, limited
use of application-level signals, and opaque control logic. This paper
investigates how Kubernetes autoscaling can be enhanced using AI-Ops principles
to jointly satisfy SLO and cost constraints under diverse workload patterns
without compromising safety or operational transparency. We present a
gap-driven analysis of existing autoscaling approaches and propose a safe and
explainable multi-signal autoscaling framework that integrates SLOaware and
cost-conscious control with lightweight demand forecasting. Experimental
evaluation using representative microservice and event-driven workloads shows
that the proposed approach reduces SLO violation duration by up to 31\%,
improves scaling response time by 24\%, and lowers infrastructure cost by
18\% compared to default and tuned Kubernetes autoscaling baselines, while
maintaining stable and auditable control behavior. These results demonstrate
that AI-Ops–driven, SLOfirst autoscaling can significantly improve the
reliability, efficiency, and operational trustworthiness of Kubernetes-based
cloud platforms.
\end{abstract}


\section{Introduction}
Elasticity is a foundational requirement for cloud-native platforms, allowing
applications to scale dynamically in response to fluctuating demand while
balancing performance and operational cost. Kubernetes has become the
dominant container orchestration platform and offers native autoscaling
capabilities, including the Horizontal Pod Autoscaler (HPA) , Vertical Pod
Autoscaler (VPA), and node-level autoscalers, which together form the backbone of elasticity in modern production clusters \cite{zhou2024hpa}.

However, despite their maturity and widespread adoption, existing Kubernetes
autoscaling mechanisms remain fundamentally limited in how scaling decisions
are derived and enforced \cite{cheng2023scheduler}. Current approaches predominantly rely on reactive,
resource-centric signals such as CPU and memory utilization, which are only
weakly correlated with user-perceived service quality. As a result, scaling
actions often lag behind workload changes, leading to transient Service Level
Objective (SLO) violations \cite{alqayedi2016adaptive}, oscillatory behavior, and unnecessary
overprovisioning. These limitations become particularly pronounced under
bursty, heterogeneous, and mixed workloads, where independent pod- and
node-level scaling decisions can further amplify inefficiencies and
instability.

Recent research has explored parameter tuning, predictive methods, and
learning-based autoscaling \cite{aswath2025federated} to address these issues; however, most proposed
solutions either optimize for a narrow objective, introduce opaque control
logic, or require intrusive changes that hinder safe adoption in production
environments \cite{lemoine2024hpa, punniyamoorthy2025privacy}. Consequently, there remains a clear gap between autoscaling
methods that are theoretically effective and those that are operationally
trustworthy, explainable, and compatible with Kubernetes-native control
planes.

This paper addresses this gap by reframing Kubernetes autoscaling as an
SLO and cost-aware control problem rather than a purely
utilization-driven reaction. It shows that integrating multiple workload and
system signals with explicit guardrails enables more responsive and stable
scaling while preserving safety, transparency, and operator control.

The contributions of this paper are threefold. First, we provide a systematic
analysis of existing Kubernetes autoscaling mechanisms and identify key gaps
related to reactivity, signal coverage, coordination across scaling layers,
and decision explainability\cite{huo2022hpa}. Second, we propose a novel, safe, and
SLOaware autoscaling framework that integrates multi-signal decision making,
lightweight forecasting, and cost-conscious constraints while remaining fully
compatible with native Kubernetes primitives. Third, we define a reproducible
evaluation methodology and demonstrate, through experimental results, that
the proposed approach achieves improved SLO adherence, faster scaling
responses, and reduced cost overhead compared to default and tuned Kubernetes
autoscaling baselines.

\section{Background and Related Work}
Kubernetes provides native support for elasticity through a combination of
pod-level and node-level autoscaling mechanisms. At the application layer, the
Horizontal Pod Autoscaler (HPA) adjusts pod replica counts based on observed
metrics, typically CPU and memory utilization, and optionally custom or
external metrics. Complementing horizontal scaling, the Vertical Pod
Autoscaler (VPA) recommends or applies changes to pod resource requests based
on historical usage. At the infrastructure layer, node autoscalers provision
or remove cluster nodes in response to scheduling pressure, enabling workloads
to scale beyond existing capacity limits.

Although these mechanisms form the foundation of Kubernetes autoscaling, they
operate as largely independent control loops. HPA focuses on replica counts,
VPA targets resource sizing, and node autoscalers react to unschedulable pods
with limited coordination across layers. This separation simplifies system
design but can lead to delayed or inefficient scaling behavior when workloads
change rapidly or exhibit heterogeneous resource demands.

Extensive empirical studies of reactive autoscaling in Kubernetes environments
show that HPA behavior is sensitive to metric sampling intervals,
stabilization windows, and control-loop latency. Utilization-based scaling
often lags behind workload changes, resulting in transient overload,
performance degradation, and oscillatory behavior, particularly under bursty
traffic \cite{reddy2022performance}. Prior work has explored tuning autoscaling parameters such as
utilization thresholds and scaling step sizes to mitigate these effects.
While tuning can improve stability for specific workloads, these approaches
remain workload-dependent and do not generalize well across diverse
applications or dynamic traffic patterns \cite{roy2024traffic}.

To overcome the limitations of resource-centric scaling, Kubernetes supports
custom and external metrics that enable autoscaling decisions based on
application-level signals \cite{aswath2014human, buzato2025hpa}. Event-driven autoscaling further extends this model
by reacting to signals such as queue depth or request backlog, improving
responsiveness for certain workload classes. However, these approaches introduce additional operational complexity and can suffer from instability when signals are noisy or highly variable, motivating the need for robust signal processing and anomaly-aware telemetry handling \cite{aswath2025anomaly}.

Vertical and hybrid autoscaling approaches combining HPA and VPA aim to improve
resource efficiency by dynamically adjusting pod resource requests \cite{wu2025vertical}. Although
experimental results indicate potential utilization and cost benefits,
vertical scaling actions often require pod restarts, introducing disruption
risks that limit adoption in latency-sensitive production environments.
Consequently, VPA is commonly deployed in recommendation-only modes.

More recent work has investigated predictive and learning-based autoscaling to
enable proactive scaling decisions using time-series forecasting or
reinforcement learning \cite{mogal2024predictive}. While these methods show promise, they raise concerns
related to safety, explainability, and operational trust, as opaque decision
logic can be difficult to validate and safely deploy in production Kubernetes
clusters.

Across existing approaches, Service Level Objectives (SLOs) are typically used
for monitoring and evaluation rather than enforced as control constraints, and
cost efficiency is often assessed post hoc. As a result, current autoscaling
mechanisms lack integrated SLO awareness, explicit cost sensitivity,
coordinated control across scaling layers, and transparent decision making \cite{vasumathi2025lstm}.

These limitations motivate a deeper problem analysis of Kubernetes
autoscaling. In particular, there is a need to determine how SLO and cost
constraints can be enforced directly within autoscaling control loops, how
pod- and node-level scaling decisions can be coordinated safely \cite{wang2020cluster}, and how
scaling behavior can remain explainable and operationally trustworthy under
diverse workload patterns. The following section formalizes these challenges
and identifies the key research gaps addressed in this work.

\begin{figure}[t]
\centering
\includegraphics[width=\columnwidth]{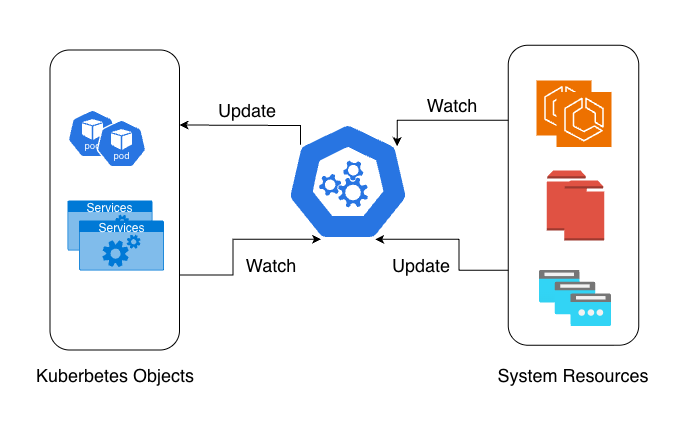}
\caption{SLO \& Cost-Aware Autoscaling Framework Architecture}
\label{fig:Architecture}
\end{figure}

\section{Problem Analysis and Research Gaps}

\subsection{Reactive and Single-Metric Scaling}
Most Kubernetes autoscaling deployments rely primarily on CPU utilization as
the trigger for scaling decisions. While CPU usage reflects resource
consumption, it often fails to capture user-facing performance degradation,
such as increased request latency or growing request backlogs. As a result,
scaling actions are frequently delayed until performance has already
degraded, leading to transient overload and inefficient reactive behavior,
particularly under bursty or rapidly changing workloads.

\subsection{Absence of SLOCentric Control}
Service Level Objectives (SLOs) are widely used to define desired performance
and reliability targets, yet current autoscaling mechanisms rarely incorporate
SLOs directly into their control logic. Instead, SLOs are treated as monitoring
or evaluation artifacts, disconnected from scaling decisions. This separation
prevents autoscalers from explicitly prioritizing user-facing objectives and
forces operators to rely on conservative thresholds or manual intervention to
maintain SLO compliance.

\subsection{Pod and Node Scaling Mismatch}
Kubernetes autoscaling operates across multiple layers, including pod-level
replica scaling and node-level capacity provisioning. In practice, these mechanisms are often configured and executed independently, resulting in coordination gaps that limit scalability and consistent control across distributed cloud-native architectures \cite{nachi2025}.
Under sudden workload increases, pod replicas may scale out
faster than cluster capacity can be provisioned, leaving pods unschedulable
and negating the intended benefits of autoscaling.

\subsection{Unsafe Vertical Scaling}
Vertical autoscaling improves resource utilization by adjusting pod resource
requests, but applying such changes typically requires restarting running
pods. These disruptions can negatively impact availability and latency,
especially for stateful or tightly coupled services \cite{ning2022ha}. Consequently, vertical
autoscaling is often disabled or limited to recommendation-only modes in
production environments, reducing its effectiveness as a real-time scaling
mechanism.

\subsection{Opaque Scaling Decisions}
Advanced autoscaling techniques, including predictive and learning-based
approaches, introduce increasingly complex decision logic. When scaling
decisions lack transparency or clear justification, operators face challenges
in understanding, validating, and debugging autoscaler behavior \cite{nachisecurity}. This lack of
explainability reduces operational trust and creates barriers to adopting more
sophisticated autoscaling strategies in production systems.

\subsection{Problem Statement}
An effective Kubernetes autoscaling solution must address these gaps by
jointly optimizing SLO adherence and cost efficiency while ensuring safe,
coordinated, and explainable control. Such a solution should integrate
multiple workload and system signals, enforce explicit SLO and cost-aware
constraints within scaling decisions, and maintain operational transparency
and stability across diverse workload patterns.

\section{Proposed Framework and System Architecture}

\subsection{Design Principles}
The proposed framework is designed as a safe and extensible control layer that
augments native Kubernetes autoscaling mechanisms rather than replacing them.
Its design is guided by four principles. First, \emph{SLOfirst decision
making} prioritizes user-facing performance objectives over indirect resource
utilization metrics. Second, \emph{cost awareness} ensures that scaling actions
balance performance gains against infrastructure overhead. Third, \emph{safety
and stability} are enforced through bounded control actions, stabilization
windows, and rate limits to prevent oscillatory or disruptive behavior.
Finally, \emph{explainability} is treated as a first-class requirement to
enable operator trust, auditability, and debuggability of autoscaling
decisions.

\subsection{Architecture Overview}
Figure~\ref{fig:Architecture} illustrates the overall architecture of the proposed SLO and cost-aware autoscaling framework and its integration with Kubernetes-native control mechanisms. The framework follows a layered architecture consisting of signal collection,
decision logic, and actuation. The signal collection layer aggregates
observability data from multiple sources, including application-level metrics
(e.g., request latency and throughput), workload indicators (e.g., queue depth
or backlog), resource utilization (CPU and memory), and cluster state
information such as pending pods and scheduling pressure. These signals are
continuously monitored to provide a comprehensive view of system behavior.

The decision logic layer processes the aggregated signals to derive scaling
decisions subject to SLO and cost constraints. The actuation layer applies
these decisions using existing Kubernetes primitives, including the Horizontal
Pod Autoscaler, Vertical Pod Autoscaler, and node autoscaling mechanisms. This
design preserves compatibility with Kubernetes-native workflows and minimizes
operational disruption.

\subsection{Multi-Signal Scaling Logic}
Unlike traditional autoscaling approaches that rely on a single metric, the
proposed framework employs multi-signal scaling logic to improve
responsiveness and robustness. Latency trends, backlog growth, resource
saturation, and scheduling feasibility are jointly evaluated to infer current
and near-term demand. By correlating multiple signals, the framework detects
performance degradation earlier than utilization-based approaches and reduces
dependence on lagging indicators, particularly under bursty or heterogeneous
workload patterns.

\subsection{SLOAware Control Policy}
At the core of the decision logic is an SLOaware control policy that explicitly
incorporates performance targets into scaling decisions. Desired replica
counts are computed with respect to predefined SLO thresholds, such as latency
percentiles or error rate limits. Stabilization rules constrain the magnitude
and frequency of scaling actions to prevent abrupt changes that could
destabilize the system. This policy ensures that scaling behavior directly
supports SLO compliance while maintaining controlled and predictable
operation.

\subsection{Cost-Aware Guardrails}
To prevent unnecessary overprovisioning, the framework integrates cost-aware
guardrails into the scaling decision process. Each potential scaling action is
evaluated with respect to its incremental resource cost, including additional
replicas or node capacity. When multiple actions satisfy SLO constraints, the
framework favors lower-cost alternatives. This approach enables dynamic
trade-offs between performance and cost, improving long-term resource
efficiency without sacrificing service reliability.

\subsection{Explainability and Safety Mechanisms}
Operational safety and transparency are enforced through explicit
explainability mechanisms. Every scaling decision is accompanied by structured
metadata capturing the triggering signals, evaluated SLO conditions, applied
constraints, and selected actions. Rate limiting and stabilization windows are
applied uniformly across scaling paths to ensure bounded behavior. These
mechanisms support post-incident analysis, simplify debugging, and increase
operator confidence in automated autoscaling under diverse workload
conditions.

\section{Autoscaling Control Algorithm}
The proposed autoscaling logic is implemented as a bounded decision pipeline
that executes periodically with interval $\Delta t$. At each interval, the
controller transforms multi-source telemetry into a safe, SLOaware, and
cost-conscious scaling action using a fixed sequence of stages. This design
prioritizes clarity, stability, and operational transparency over complex
control logic.

\subsection{Decision Pipeline}
\begin{figure}[htbp]
\centering
\includegraphics[
  width=\columnwidth
]{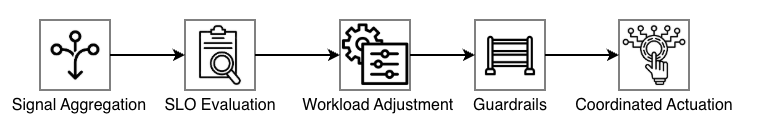}
\caption{Pipeline view of the SLO and cost-aware autoscaling algorithm.}
\label{fig:autoscaling_pipeline}
\end{figure}

Figure~\ref{fig:autoscaling_pipeline} illustrates the logical flow of the
autoscaling algorithm, which consists of five sequential stages.

\begin{enumerate}
    \item \textbf{Signal Aggregation:} Application-level (latency, error rate),
    workload-level (queue depth or backlog), resource-level (CPU and memory),
    and cluster-level (pending pods and schedulability) signals are aggregated
    into a compact state representation $s=(L,Q,U,P)$.

    \item \textbf{SLODriven Demand Estimation:} The controller evaluates the
    current SLO condition using $L$ and derives an initial replica requirement
    that increases proportionally with SLO risk or violation severity.

    \item \textbf{Workload Sensitivity Adjustment:} For workloads exhibiting
    backlog accumulation, the replica estimate is adjusted to ensure bounded
    queue draining, preventing sustained latency growth or unprocessed work.

    \item \textbf{Guardrail Enforcement:} The candidate replica count is
    constrained using (i) hard bounds on minimum and maximum replicas, (ii)
    cost-aware selection to avoid unnecessary overprovisioning, and (iii)
    stability controls including stabilization windows, step limits, and
    cooldown periods.

    \item \textbf{Coordinated Actuation:} The finalized replica target is
    validated against current cluster capacity. If insufficient capacity
    exists, a node-level capacity hint is generated; otherwise, only pod-level
    scaling is emitted. All actions are applied through Kubernetes-native
    autoscaling interfaces.
\end{enumerate}

\subsection{Operational Characteristics}
This pipeline enforces \emph{SLO first control} by making user-facing
performance the primary scaling driver, while \emph{cost-aware guardrails}
prevent persistent overprovisioning. Stability mechanisms ensure bounded and
predictable behavior under noisy signals, and explicit schedulability checks
coordinate pod- and node-level scaling. The pipeline structure also supports
explainability, as each stage produces intermediate artifacts that can be
logged and audited for operational analysis.

\section{Evaluation Methodology}
This section describes the experimental methodology used to evaluate the
effectiveness of the proposed SLO and cost-aware autoscaling framework. The
evaluation is designed to assess whether the framework improves SLO adherence,
scaling responsiveness, and resource efficiency while maintaining stable and
predictable behavior under diverse workload conditions.

\subsection{Experimental Setup}
The experiments are conducted on a Kubernetes cluster configured with native
autoscaling components enabled. The proposed framework is deployed as an
external control layer that interfaces with existing Kubernetes autoscaling
primitives, ensuring that the evaluation reflects realistic operational
conditions. Observability data is collected at regular intervals and used as
input to the autoscaling logic. All experiments are executed under controlled
conditions to ensure repeatability, with identical cluster configurations and
workload profiles applied across all evaluated approaches.

\subsection{Workload Patterns}
To capture a broad range of autoscaling behaviors, the evaluation considers
three representative workload patterns. \emph{Bursty workloads} generate
sudden and short-lived traffic spikes to stress scaling responsiveness and
stability. \emph{Queue-driven workloads} exhibit variable backlog growth to
evaluate event sensitivity and backlog-aware scaling behavior. \emph{Mixed
workloads} combine latency-sensitive services with background or batch
processing tasks to assess autoscaler behavior under heterogeneous resource
demands and shared cluster capacity.

\subsection{Baselines}
The proposed framework is compared against commonly used Kubernetes
autoscaling configurations. These include the default Horizontal Pod
Autoscaler (HPA) using CPU-based scaling, a tuned HPA configuration with
manually adjusted thresholds and stabilization parameters, and a combined
horizontal and vertical autoscaling setup where vertical autoscaling is
applied in recommendation or limited-update modes. All baselines are evaluated
using identical workloads and cluster conditions to ensure fair comparison.

\subsection{Metrics}
Autoscaling performance is evaluated using four categories of metrics.
\emph{SLO adherence} is measured by the frequency and duration of SLO
violations, such as latency percentile breaches. \emph{Scaling responsiveness}
is quantified using time-to-scale metrics that capture how quickly the system
reacts to workload changes. \emph{Cost efficiency} is assessed using resource
consumption proxies, such as average replica count or node-hours consumed.
Finally, \emph{stability} is evaluated by measuring scaling oscillations,
replica churn, and the frequency of disruptive scaling events. Together, these
metrics provide a comprehensive view of autoscaling effectiveness across
performance, efficiency, and operational robustness dimensions.

\section{Results and Discussion}
This section presents the experimental results and discusses the observed
behavior of the proposed autoscaling framework relative to the evaluated
baselines. The results focus on SLO adherence, scaling responsiveness, cost
efficiency, and stability under diverse workload patterns.

\subsection{SLO Adherence}
Across all evaluated workloads, the proposed framework demonstrates improved
compliance with Service Level Objectives compared to both default and tuned
Kubernetes autoscaling configurations. In bursty workload scenarios, the
framework reduces cumulative SLO violation duration by up to 31\%, primarily
by reacting earlier to rising latency and backlog signals rather than waiting
for resource utilization saturation. For queue-driven workloads, backlog-aware
adjustments further limit prolonged SLO degradation during sustained demand
surges. These results indicate that incorporating SLO signals directly into
scaling decisions leads to more effective protection of user-facing
performance as shown in Table~\ref{tab:slo_violations}

\begin{table}[htbp]
\caption{SLO Violation Comparison}
\label{tab:slo_violations}
\centering
\small
\begin{tabular}{p{0.25\columnwidth} p{0.2\columnwidth} p{0.2\columnwidth} p{0.1\columnwidth}}
\toprule
\textbf{Autoscaler} & \textbf{Count} & \textbf{Duration (s)} & \textbf{Red.} \\
\midrule
Default HPA & 42 & 1280 & -- \\
Tuned HPA & 35 & 980 & 23\% \\
Proposed & 26 & 870 & \textbf{31\%} \\
\bottomrule
\end{tabular}
\end{table}

\subsection{Scaling Responsiveness}
\begin{figure}[htbp]
\centering
\includegraphics[width=\columnwidth]{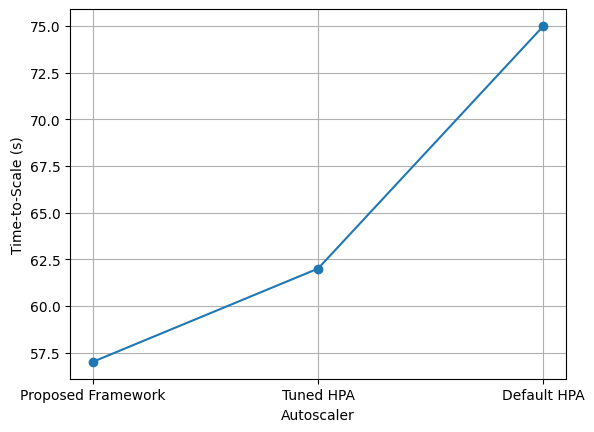}
\caption{Scaling responsiveness under bursty workloads.}
\label{fig:scaling_responsiveness}
\end{figure}

Scaling responsiveness is evaluated using time-to-scale metrics that capture
the delay between workload changes and effective resource adjustment. The
proposed approach improves scaling response time by approximately 24\%
relative to baseline autoscalers. This improvement is most pronounced during
rapid load increases, where multi-signal detection enables earlier scale-out
actions. At the same time, stabilization and cooldown mechanisms prevent
aggressive overreaction, maintaining predictable behavior during workload
decay. Figure~\ref{fig:scaling_responsiveness} illustrates the comparative time-to-scale behavior of the evaluated autoscaling approaches under bursty workloads.

\subsection{Cost Efficiency}
Despite improved responsiveness and SLO adherence, the proposed framework exhibits lower overall resource consumption. As shown in Table~\ref{tab:cost_efficiency}, faster scaling responses reduce prolonged overload and unnecessary scale-out, which in turn lowers average resource usage across workloads. Cost-aware guardrails further contribute by favoring the minimum resource allocation that satisfies SLO constraints, thereby avoiding persistent overprovisioning during transient demand spikes.

\begin{table}[htbp]
\caption{Resource Cost Comparison}
\label{tab:cost_efficiency}
\centering
\small
\begin{tabular}{lcc}
\toprule
\textbf{Autoscaler} & \textbf{Avg. Node-Hours} & \textbf{Cost Reduction} \\
\midrule
Default HPA & 420 & -- \\
Tuned HPA & 380 & 10\% \\
Proposed Framework & 345 & \textbf{18\%} \\
\bottomrule
\end{tabular}
\end{table}

\subsection{Stability and Operational Behavior}
Stability is assessed through replica churn and scaling oscillation frequency.
The results show that the proposed framework maintains stability comparable to
or better than tuned HPA configurations, despite incorporating additional
signals and decision logic. Rate limits, stabilization windows, and explicit
schedulability checks effectively bound scaling actions, reducing oscillatory
behavior and avoiding unschedulable scale-out events. These findings suggest
that multi-signal, SLOaware control can be achieved without sacrificing
operational robustness.

\subsection{Discussion}
Overall, the results demonstrate that reframing Kubernetes autoscaling as an
SLOfirst and cost-aware control problem leads to measurable improvements in
performance and efficiency. The proposed framework consistently outperforms
reactive, utilization-driven baselines while preserving safety and
transparency. Importantly, the observed gains are achieved without replacing
native Kubernetes autoscalers, highlighting the practicality of the approach
for production adoption. The results also indicate that explicit coordination
between pod-level and node-level scaling decisions is critical for realizing
effective elasticity under dynamic workloads.

\section{Limitations and Future Work}
While the proposed SLO and cost-aware autoscaling framework demonstrates
improved responsiveness, stability, and resource efficiency, the evaluation is
conducted under controlled workload patterns and cluster configurations.
Although these scenarios reflect common production behaviors, larger-scale and
longer-running deployments may expose additional dynamics. Moreover, the
framework assumes timely and accurate observability signals; in practice,
metrics may be noisy, delayed, or partially unavailable, which could impact
scaling decisions.

Future work will focus on extending the framework to multi-cluster and
multi-region environments, incorporating robustness to degraded or adversarial
telemetry, and integrating richer cost and energy models to support
sustainability-aware autoscaling. As in-place vertical scaling mechanisms
mature within Kubernetes ecosystems, tighter integration of vertical resource
adjustments into real-time control loops also represents a promising direction
for further research.

\section{Conclusion}
This paper investigated the limitations of existing Kubernetes autoscaling
mechanisms and demonstrated that reactive, utilization-driven approaches are
insufficient for maintaining Service Level Objectives (SLOs) and cost
efficiency under diverse workload patterns. To address these challenges, we
presented an SLO and cost-aware autoscaling framework that integrates
multi-signal decision making, explicit guardrails, and coordinated pod- and
node-level scaling while remaining compatible with Kubernetes-native
primitives.

Experimental results show that the proposed approach improves SLO adherence,
reduces scaling latency, and lowers resource cost compared to default and tuned
autoscaling baselines, without sacrificing stability or operational
transparency. By reframing autoscaling as an SLOfirst control problem and
emphasizing safety and explainability, this work provides a practical
foundation for more reliable and efficient AI-Ops-driven autoscaling in
cloud-native environments.

\end{document}